\RequirePackage[left]{lineno}
\documentclass[aps,prl,twocolumn,superscriptaddress,showpacs,preprintnumbers,amsmath,amssymb]{revtex4}
\usepackage{graphicx} 
\usepackage{dcolumn}  
\usepackage{lineno, blindtext}
\usepackage{color}
\usepackage[normalem]{ulem} 

\newcommand{\BSTBST}{\mbox{$B_{s}^{*}\bar B_{s}^{*}$}}
\newcommand{\BSTBS}{\mbox{$B_{s}^{*}\bar B_{s}$}}
\newcommand{\BSBS}{\mbox{$B_{s}\bar B_{s}$}}

\graphicspath{{ps}}

\begin{document}


\preprint{\vbox{ \hbox{   }
    \hbox{Belle Preprint 2021-26}
          		         \hbox{KEK Preprint 2021-31}
}}

\title{ \quad\\[1.0cm] Search for the decay {\boldmath$B_s^0\rightarrow\eta\eta$}}



\noaffiliation
\affiliation{University of the Basque Country UPV/EHU, 48080 Bilbao}
\affiliation{Beihang University, Beijing 100191}
\affiliation{Brookhaven National Laboratory, Upton, New York 11973}
\affiliation{Budker Institute of Nuclear Physics SB RAS, Novosibirsk 630090}
\affiliation{Faculty of Mathematics and Physics, Charles University, 121 16 Prague}
\affiliation{Chonnam National University, Kwangju 660-701}
\affiliation{University of Cincinnati, Cincinnati, Ohio 45221}
\affiliation{Deutsches Elektronen--Synchrotron, 22607 Hamburg}
\affiliation{Duke University, Durham, North Carolina 27708}
\affiliation{Key Laboratory of Nuclear Physics and Ion-beam Application (MOE) and Institute of Modern Physics, Fudan University, Shanghai 200443}
\affiliation{Justus-Liebig-Universit\"at Gie\ss{}en, 35392 Gie\ss{}en}
\affiliation{SOKENDAI (The Graduate University for Advanced Studies), Hayama 240-0193}
\affiliation{Gyeongsang National University, Chinju 660-701}
\affiliation{Hanyang University, Seoul 133-791}
\affiliation{University of Hawaii, Honolulu, Hawaii 96822}
\affiliation{High Energy Accelerator Research Organization (KEK), Tsukuba 305-0801}
\affiliation{J-PARC Branch, KEK Theory Center, High Energy Accelerator Research Organization (KEK), Tsukuba 305-0801}
\affiliation{Forschungszentrum J\"{u}lich, 52425 J\"{u}lich}
\affiliation{IKERBASQUE, Basque Foundation for Science, 48013 Bilbao}
\affiliation{Indian Institute of Science Education and Research Mohali, SAS Nagar, 140306}
\affiliation{Indian Institute of Technology Bhubaneswar, Satya Nagar 751007}
\affiliation{Indian Institute of Technology Guwahati, Assam 781039}
\affiliation{Indian Institute of Technology Hyderabad, Telangana 502285}
\affiliation{Indian Institute of Technology Madras, Chennai 600036}
\affiliation{Indiana University, Bloomington, Indiana 47408}
\affiliation{Institute of High Energy Physics, Chinese Academy of Sciences, Beijing 100049}
\affiliation{Institute of High Energy Physics, Vienna 1050}
\affiliation{INFN - Sezione di Napoli, 80126 Napoli}
\affiliation{INFN - Sezione di Torino, 10125 Torino}
\affiliation{Advanced Science Research Center, Japan Atomic Energy Agency, Naka 319-1195}
\affiliation{J. Stefan Institute, 1000 Ljubljana}
\affiliation{Institut f\"ur Experimentelle Teilchenphysik, Karlsruher Institut f\"ur Technologie, 76131 Karlsruhe}
\affiliation{Kennesaw State University, Kennesaw, Georgia 30144}
\affiliation{King Abdulaziz City for Science and Technology, Riyadh 11442}
\affiliation{Department of Physics, Faculty of Science, King Abdulaziz University, Jeddah 21589}
\affiliation{Kitasato University, Sagamihara 252-0373}
\affiliation{Korea Institute of Science and Technology Information, Daejeon 305-806}
\affiliation{Korea University, Seoul 136-713}
\affiliation{Kyoto University, Kyoto 606-8502}
\affiliation{Kyungpook National University, Daegu 702-701}
\affiliation{LAL, Univ. Paris-Sud, CNRS/IN2P3, Universit\'{e} Paris-Saclay, Orsay}
\affiliation{\'Ecole Polytechnique F\'ed\'erale de Lausanne (EPFL), Lausanne 1015}
\affiliation{P.N. Lebedev Physical Institute of the Russian Academy of Sciences, Moscow 119991}
\affiliation{Liaoning Normal University, Dalian 116029}
\affiliation{Faculty of Mathematics and Physics, University of Ljubljana, 1000 Ljubljana}
\affiliation{Ludwig Maximilians University, 80539 Munich}
\affiliation{Luther College, Decorah, Iowa 52101}
\affiliation{University of Malaya, 50603 Kuala Lumpur}
\affiliation{University of Maribor, 2000 Maribor}
\affiliation{Max-Planck-Institut f\"ur Physik, 80805 M\"unchen}
\affiliation{School of Physics, University of Melbourne, Victoria 3010}
\affiliation{University of Mississippi, University, Mississippi 38677}
\affiliation{University of Miyazaki, Miyazaki 889-2192}
\affiliation{Moscow Physical Engineering Institute, Moscow 115409}
\affiliation{Moscow Institute of Physics and Technology, Moscow Region 141700}
\affiliation{Graduate School of Science, Nagoya University, Nagoya 464-8602}
\affiliation{Kobayashi-Maskawa Institute, Nagoya University, Nagoya 464-8602}
\affiliation{Universit\`{a} di Napoli Federico II, 80055 Napoli}
\affiliation{Nara Women's University, Nara 630-8506}
\affiliation{National Central University, Chung-li 32054}
\affiliation{National United University, Miao Li 36003}
\affiliation{Department of Physics, National Taiwan University, Taipei 10617}
\affiliation{H. Niewodniczanski Institute of Nuclear Physics, Krakow 31-342}
\affiliation{Nippon Dental University, Niigata 951-8580}
\affiliation{Niigata University, Niigata 950-2181}
\affiliation{Novosibirsk State University, Novosibirsk 630090}
\affiliation{Osaka City University, Osaka 558-8585}
\affiliation{Pacific Northwest National Laboratory, Richland, Washington 99352}
\affiliation{Peking University, Beijing 100871}
\affiliation{University of Pittsburgh, Pittsburgh, Pennsylvania 15260}
\affiliation{Punjab Agricultural University, Ludhiana 141004}
\affiliation{Research Center for Nuclear Physics, Osaka University, Osaka 567-0047}
\affiliation{Theoretical Research Division, Nishina Center, RIKEN, Saitama 351-0198}
\affiliation{University of Science and Technology of China, Hefei 230026}
\affiliation{Seoul National University, Seoul 151-742}
\affiliation{Showa Pharmaceutical University, Tokyo 194-8543}
\affiliation{Soongsil University, Seoul 156-743}
\affiliation{University of South Carolina, Columbia, South Carolina 29208}
\affiliation{Sungkyunkwan University, Suwon 440-746}
\affiliation{School of Physics, University of Sydney, New South Wales 2006}
\affiliation{Department of Physics, Faculty of Science, University of Tabuk, Tabuk 71451}
\affiliation{Tata Institute of Fundamental Research, Mumbai 400005}
\affiliation{Department of Physics, Technische Universit\"at M\"unchen, 85748 Garching}
\affiliation{Toho University, Funabashi 274-8510}
\affiliation{Department of Physics, Tohoku University, Sendai 980-8578}
\affiliation{Earthquake Research Institute, University of Tokyo, Tokyo 113-0032}
\affiliation{Department of Physics, University of Tokyo, Tokyo 113-0033}
\affiliation{Tokyo Institute of Technology, Tokyo 152-8550}
\affiliation{Tokyo Metropolitan University, Tokyo 192-0397}
\affiliation{Virginia Polytechnic Institute and State University, Blacksburg, Virginia 24061}
\affiliation{Wayne State University, Detroit, Michigan 48202}
\affiliation{Yamagata University, Yamagata 990-8560}
\affiliation{Yonsei University, Seoul 120-749}

  \author{B.~Bhuyan}\affiliation{Indian Institute of Technology Guwahati, Assam 781039} 
  \author{K.~J.~Nath}\affiliation{Indian Institute of Technology Guwahati, Assam 781039} 
  \author{J.~Borah}\affiliation{Indian Institute of Technology Guwahati, Assam 781039} 
  \author{I.~Adachi}\affiliation{High Energy Accelerator Research Organization (KEK), Tsukuba 305-0801}\affiliation{SOKENDAI (The Graduate University for Advanced Studies), Hayama 240-0193} 
  \author{H.~Aihara}\affiliation{Department of Physics, University of Tokyo, Tokyo 113-0033} 
  \author{S.~Al~Said}\affiliation{Department of Physics, Faculty of Science, University of Tabuk, Tabuk 71451}\affiliation{Department of Physics, Faculty of Science, King Abdulaziz University, Jeddah 21589} 
  \author{D.~M.~Asner}\affiliation{Brookhaven National Laboratory, Upton, New York 11973} 
  \author{H.~Atmacan}\affiliation{University of South Carolina, Columbia, South Carolina 29208} 
  \author{V.~Aulchenko}\affiliation{Budker Institute of Nuclear Physics SB RAS, Novosibirsk 630090}\affiliation{Novosibirsk State University, Novosibirsk 630090} 
  \author{T.~Aushev}\affiliation{Moscow Institute of Physics and Technology, Moscow Region 141700} 
% ...
\collaboration{The Belle Collaboration}

\begin{abstract}
We report results from a search for the decay $B_s^0\rightarrow\eta\eta$ using 121.4 fb$^{-1}$ of data collected at the $\Upsilon(5S)$ resonance with the Belle detector at the KEKB asymmetric-energy $e^+e^-$ collider. We do not observe any signal and set an upper limit on the branching fraction of $14.3\times 10^{-5}$ at $90\%$ confidence level. This result represents a significant improvement over the previous most stringent limit.

\end{abstract}


\maketitle

\tighten

{\renewcommand{\thefootnote}{\fnsymbol{footnote}}}
\setcounter{footnote}{0}

Studies of two-body charmless $B$ decays can shed light on the validity of the various theory approaches which are used to study them. The decay  $B_s^0\rightarrow\eta\eta$ is a neutral charmless process that can occur through a variety of amplitudes such as the Cabibbo-suppressed $b\rightarrow u$ transition or one loop diagrams with a quark and a virtual $W^{\pm}$ boson, as shown in Fig. 1. Contributions can also come from electroweak ``penguin" diagrams. Theoretically, the decay has been studied within the framework of soft-collinear effective theory \cite{scet}, perturbative quantum chromodynamics (QCD) \cite{pqcd}, and QCD factorization \cite{qcdf}. All these approaches predict a branching fraction ($\cal{B}$) in the range of $(7-16)\times 10^{-6}$, albeit with large model uncertainties. Thus far, the only experimental result is an upper limit (UL) on $\mathcal{B}$ at $90\%$ confidence level (CL) of $1.5\times 10^{-3}$, obtained by the L3 experiment \cite{l3}. This analysis constitutes the first attempt to search for this decay using $e^{+}e^{-}$ collision data recorded by the Belle experiment.

%
%


\begin{figure}[htb]
\centering
\includegraphics[height=5cm]{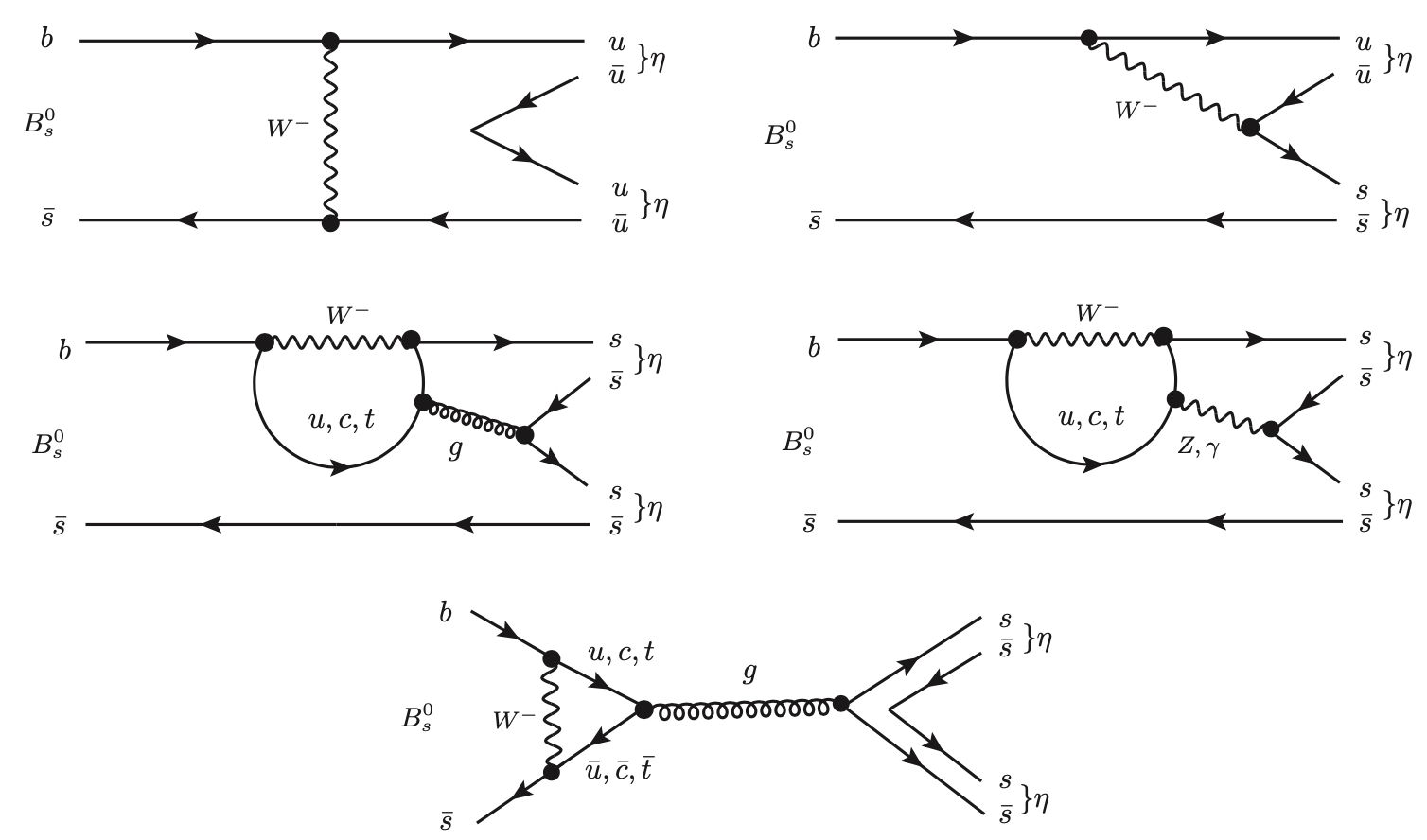}
\caption{Feynman diagrams for $B_s^0\rightarrow\eta\eta$. }
\end{figure}

The Belle detector \cite{detector1, detector2} is a large-solid-angle magnetic spectrometer that consists of a silicon vertex detector, a 50-layer central drift chamber (CDC), an array of aerogel threshold Cherenkov counters (ACC), a barrel-like arrangement of time-of-flight scintillation counters (TOF), and an electromagnetic calorimeter composed of CsI(TI) crystals (ECL). All these components are located inside a superconducting solenoid coil that provides a 1.5 T magnetic field. The analysis is based on $121.4$ fb$^{-1}$ of data collected by Belle near the $\Upsilon(5S)$ resonance, corresponding to $(16.60\pm 2.68)\times 10^6$ $B_s^0$ mesons produced.


The $b\bar{b}$ production cross section at the $\Upsilon(5S)$ center-of-mass energy is measured to be $\sigma_{b\bar{b}}^{\Upsilon(5S)} = (0.340 \pm 0.016)$ nb~\cite{Sevda}, while the fraction of $B_{s}^{(*)}\bar{B}_{s}^{(*)}$ in $b\bar{b}$ events is $f_{s} = 0.201 \pm 0.031$~\cite{pdg2020}. The $B_{s}^{(*)}\bar{B}_{s}^{(*)}$ pairs include $\BSTBST$, $\BSTBS$, and $\BSBS$, the fraction of the first two being $f_{\tiny{\BSTBST}} = (87.0 \pm 1.7)\%$ and $f_{\tiny{\BSTBS}} = (7.3 \pm 1.4)\%$, respectively ~\cite{Sevda}. The $B_{s}^{*}$ mesons decay to ground-state $B_{s}^{0}$ mesons via the emission of a photon. 

To reconstruct $B_s^0\rightarrow\eta\eta$ candidates, we first reconstruct $\eta$ candidates from either  $\eta\rightarrow\gamma\gamma$ ($\eta_{\gamma\gamma}$) or $\eta\rightarrow\pi^+\pi^-\pi^0$ ($\eta_{3\pi}$) decays. To calculate the experimental acceptance and reconstruction efficiency, signal Monte Carlo (MC) events are generated for the $B_s^{0}\rightarrow\eta_{\gamma\gamma}\eta_{\gamma\gamma} $, $B_s^{0}\rightarrow\eta_{\gamma\gamma}\eta_{3\pi}$ and $B_s^{0}\rightarrow\eta_{3\pi}\eta_{3\pi}$ modes with the EvtGen \cite{evtgen} event generator. Backgrounds are due to the copious production of quark-antiquark ($u,d,s,c$) pairs in $e^+e^-$ annihilation; this is referred to as the continuum background. Additional background arises from $B_s^{(*)}\bar B_s^{(*)}$ decays (referred to as $bsbs$) and $B^*\bar{B}^*$, $B^*\bar{B}$, $B\bar{B}$, $B^{*}\bar B^{*}\pi$, $B^*\bar{B}\pi$, $B\bar{B}\pi$ and $B\bar{B}\pi\pi$ decays (referred to as $nonbsbs$) from $B^0$ and $B^{\pm}$ near $\Upsilon(5S)$ resonance. Dedicated MC samples are generated to study these background processes. The detector response is simulated using GEANT3 \cite{geant3}, with beam-related backgrounds from data being embedded to produce a more realistic simulated event sample.



Photons are reconstructed by identifying energy deposits in the ECL that are not associated with any charged track. We require a minimum energy of $0.1$ GeV. The timing characteristics of energy clusters used in photon reconstruction must be consistent with the beam collision time, which is determined at the trigger level for the candidate event. Daughter photons from $\pi^0$ decays pose a significant background for $\eta\rightarrow\gamma\gamma$ reconstruction. To suppress this background, we calculate a $\pi^0$ likelihood for each pair of photons using the energies and polar angles of the photons as well as the diphoton invariant mass. This likelihood is optimized using the figure-of-merit, $S/\sqrt{S + B}$ to increase the signal significance, where $S$ and $B$ represents the estimated number of signal and background events, respectively. The optimized value of the likelihood is $0.4$, which is $90\%$ efficient in selecting the signal and rejects $62\%$ of backgrounds.

To reject merged $\gamma$'s and neutral hadrons, the ratio of the energy deposited in an innermost $(3\times3)$ array of crystals compared to that deposited in the $(5\times5)$ array centered around the most energetic crystal is required to be greater than $0.95$. Selected photons are combined to form $ \eta_{\gamma\gamma} $ candidates, with the diphoton invariant mass required to be in the range $(0.50 - 0.60)$ ~GeV/${\it c}^2$. This range corresponds to $\pm 3\sigma$ in resolution around the $\eta$ mass. The candidate $\pi^0$ is required to have a $\gamma\gamma$ invariant mass between 0.117~GeV/${\it c}^2$  and 0.149 GeV/${\it c}^2$, which corresponds to $\pm 3\sigma$ in resolution around the nominal $\pi^0$ mass. For the reconstruction of both $\eta$ and $\pi^0$ candidates, we perform  mass-constrained fits to improve the momentum resolution.





Charged tracks must have an impact parameter with respect to the interaction point of less than $0.3$ $\rm cm$ in the $\rm r - \mathrm  \phi$ plane, and less than 3.0 $\rm cm$ along the $e^+$ beam direction. The $\rm r - \mathrm  \phi$ plane is perpendicular to the $e^+$ beam direction, where, $\rm r$ represents the radius of the hit and $\mathrm \phi$ measures the azimuthal angle. The transverse momentum of the selected tracks  is required to be greater than $0.1$ GeV/$\it c$. Charged pions are identified using information obtained from the CDC, the TOF, and the ACC. This information is combined to form a likelihood $(\mathcal{L})$ for hadron identification. We require that charged pions satisfy $\mathcal{L}_{K}/(\mathcal{L}_{K}+\mathcal{L}_{\pi}) < 0.4$, where $\mathcal{L}_{K} (\mathcal{L}_{\pi})$ denotes the likelihood for a track to be a kaon (pion). The efficiency of this requirement is approximately $92\%$, while the probability for a kaon to be misidentified as a pion is about $8\%$. Two oppositely charged pions are combined with a $\pi^0$ candidate to reconstruct an $\eta_{3\pi}$ candidate, with the resulting invariant mass required to lie in the range $(0.527-0.568)$ GeV/$c^{\it 2}$. This range corresponds to $\pm 3\sigma$ in resolution around the $\eta$ mass. For each such $\eta$ candidate, a mass-constrained fit is performed and the $\chi^2$ is required to be less than $20$, to reject backgrounds from low energy photons.

Candidate $B_s^{0}\rightarrow\eta\eta$ decays are formed by combining a pair of $\eta$ mesons, and further selections are applied to their beam-energy constrained mass ($M_{\rm bc}$) and the energy difference ($\Delta E$). These quantities are defined as

\begin{linenomath}
\begin{equation}
M_{\rm bc}=\frac{\sqrt{{(E_{\rm beam})^{2}-(\vec{p}_{\rm reco})^{2}c^2}}}{c^2}
\end{equation}
\end{linenomath}

\begin{linenomath}
\begin{equation}
\Delta E=E_{\rm reco}-E_{\rm beam},
\end{equation}
\end{linenomath}

where $E_{\rm beam}$ is the beam energy, and $\vec{p}_{\rm reco}$ and $E_{\rm reco}$ are the momentum and energy, respectively, of the reconstructed $B_s^0$ candidate. All quantities are calculated in the $e^+e^-$ center-of-mass frame. Signal candidates are required to satisfy 5.30 GeV/${\it c}^2$ $<M_{\rm bc}<5.44$ GeV/${\it c}^2$ and $-0.60$ GeV$<\Delta E<0.20$ GeV.

The dominant source of background is continuum events. As the outgoing light quarks carry significant momenta, these events tend to be jet-like and thus topologically different from more  spherical $B_{s}^{(*)}\bar{B}_{s}^{(*)}$ events, in which the $B^0_s$ mesons have small momenta. To suppress this background, a neural network (NN) based on the NeuroBayes algorithm \cite{nn} is used. The inputs consist of sixteen event shape variables that include modified Fox-Wolfram moments~\cite{ksfw}, and the absolute value of the cosine of the angle between the thrust axis \cite{Bfactories} of the $B_{s}^0$ decay products and the rest of the event. The NN output, $\cal{C}_{\mathrm{NN}}$, ranges from $-1$ to $+1$, with a value near +1 $(\mathrm{-1})$ being more likely due to a signal (background) event.



To reduce the continuum background, $\cal{C}_{\mathrm{NN}}$ is required to be greater than $-0.6$. This requirement suppresses the continuum background by a factor of $2.5$, with a signal efficiency of about $95\%$. The contribution from other background processes such as $bsbs$ and $nonbsbs$ is estimated to be less than one event. The ${\cal C}_{\rm NN}$ value is transformed to a new variable, $\cal{C}'_{\mathrm{NN}}$, to facilitate its modeling with a simple analytical function:

\begin{linenomath}
\begin{equation}
\cal{C}'_{\mathrm{NN}} = \log \bigg(\frac{\cal{C}_{\mathrm{NN}} - \cal{C}_{\mathrm{NN(min)}}}{\cal{C}_{\mathrm{NN(max)}} - \cal{C}_{\mathrm{NN}}}\bigg),
\end{equation}
\end{linenomath}

where $\cal{C}_{\mathrm{NN(min)}}$ =$-0.6$ and $\cal{C}_{\mathrm{NN(max)}}$ is the maximum value of $\cal{C}_{\mathrm{NN}}$ obtained from the NN distribution.


After applying all the selection criteria, about $6\%$ of events in the $\eta_{\gamma\gamma}\eta_{\gamma\gamma}$  decay mode are found to have more than one signal candidate, whereas for the $\eta_{\gamma\gamma}\eta_{3\pi}$ and $\eta_{3\pi}\eta_{3\pi}$ decay modes, about $16\%$ and $28\%$ of events, respectively, have more than one signal candidate. For events with multiple candidates, we select the  candidate having the smallest value of the sum of $\chi^2$ values from the two mass-constrained fits. The overall efficiency of the best candidate selection criteria obtained from a signal MC study is $94\%$. The fraction of mis-reconstructed signal events, referred to as self-cross-feed (SCF), is found from MC studies to be $ 5.5\% $, $ 8.9\% $, and $12.9\%$ for the $\eta_{\gamma\gamma}\eta_{\gamma\gamma}$, $\eta_{\gamma\gamma}\eta_{3\pi}$, and $\eta_{3\pi}\eta_{3\pi}$ decay modes, respectively.

\begin{table}
\caption{\label{tab:table1} PDFs used to model the $M_{\rm bc}$, $\Delta E$, and $\cal{C}'_{\mathrm{NN}}$ distributions. The notations G, CB, CP, and A correspond to Gaussian, Crystal Ball ~\cite{CBall}, Chebyshev polynomial, and ARGUS functions ~\cite{Argus}, respectively.}
\begin{ruledtabular}
\begin{tabular}{ c c c c}

Fit component & $M_{\rm bc}$ & $\Delta E$ & $\cal{C}'_{\mathrm{NN}}$ \\
\hline
\small{
Signal} & CB & CB + G & G + G\\
\hline
SCF & CB & CB + CP & G + G\\
\hline
Continuum & A & CP & G \\

\end{tabular}
\end{ruledtabular}

\end{table}

\begin{figure*}
\centering
\includegraphics[ height=4cm, width=5.9cm]{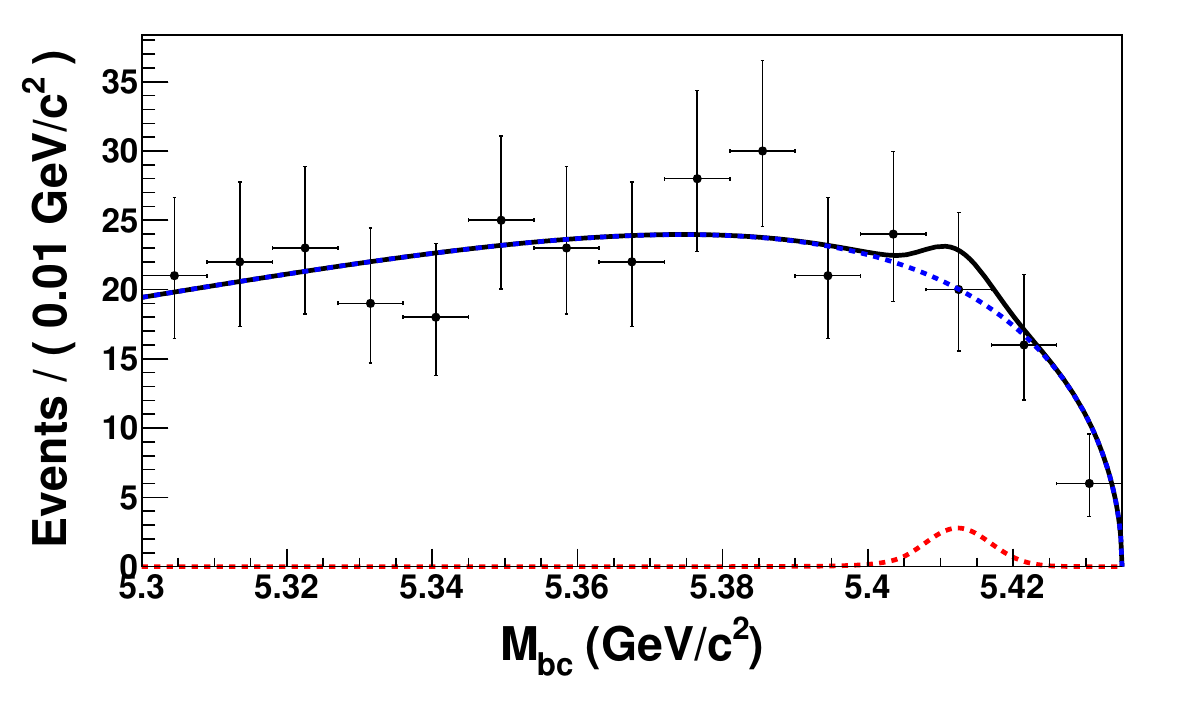}
\includegraphics[ height=4cm, width=5.9cm]{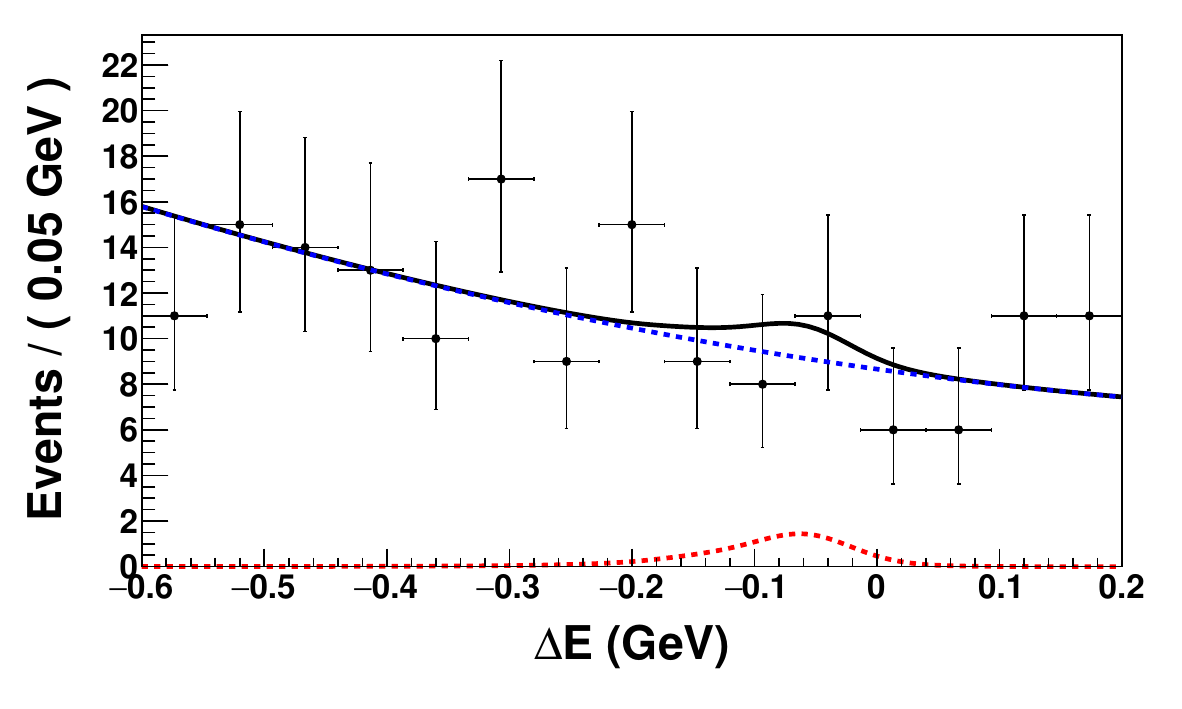}
\includegraphics[ height=4cm, width=5.9cm]{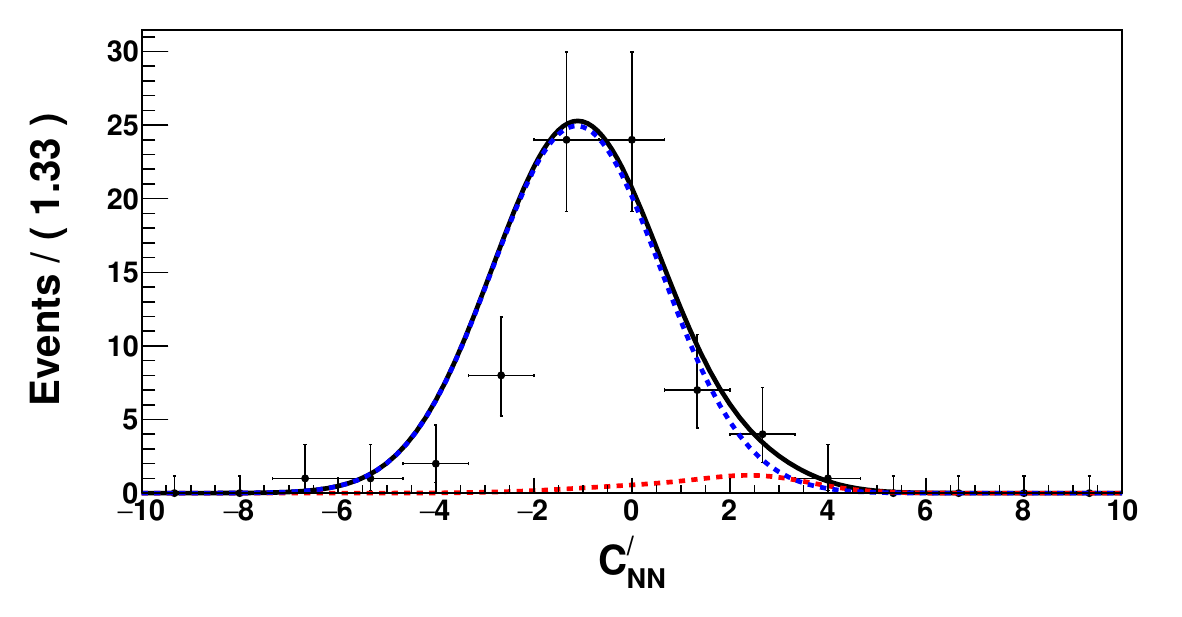}
\label{data_fit_22}
\centering
\includegraphics[ height=4cm, width=5.9cm]{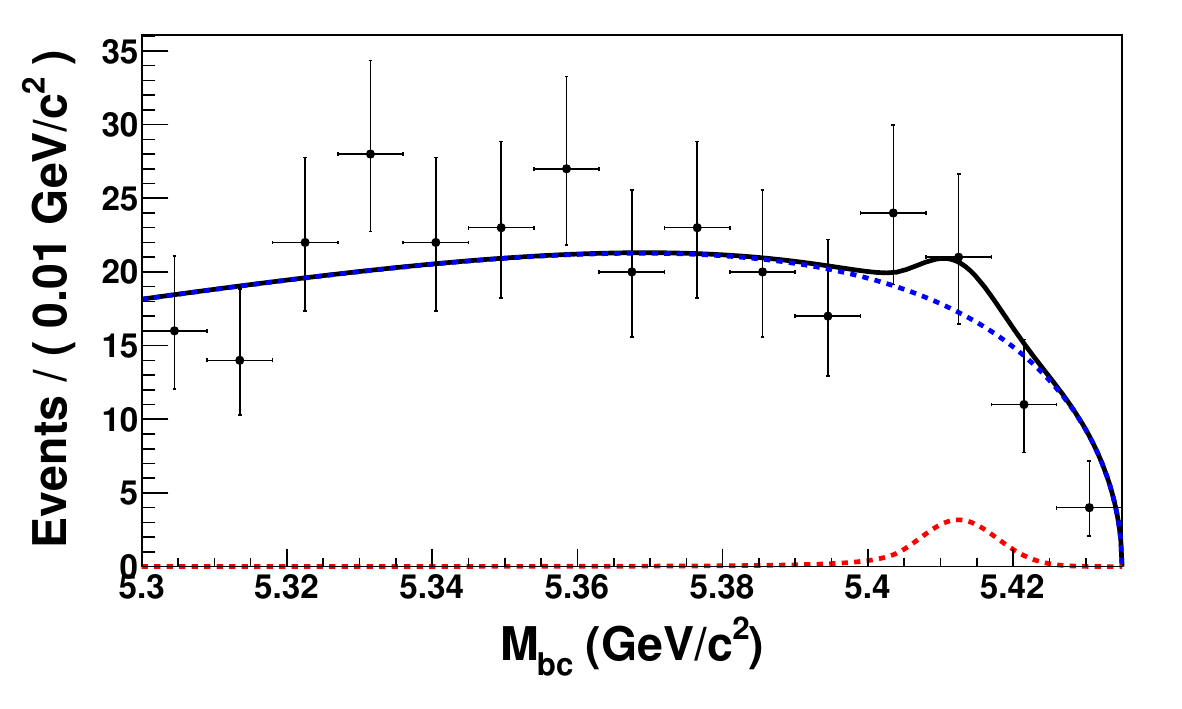}
\includegraphics[ height=4cm, width=5.9cm]{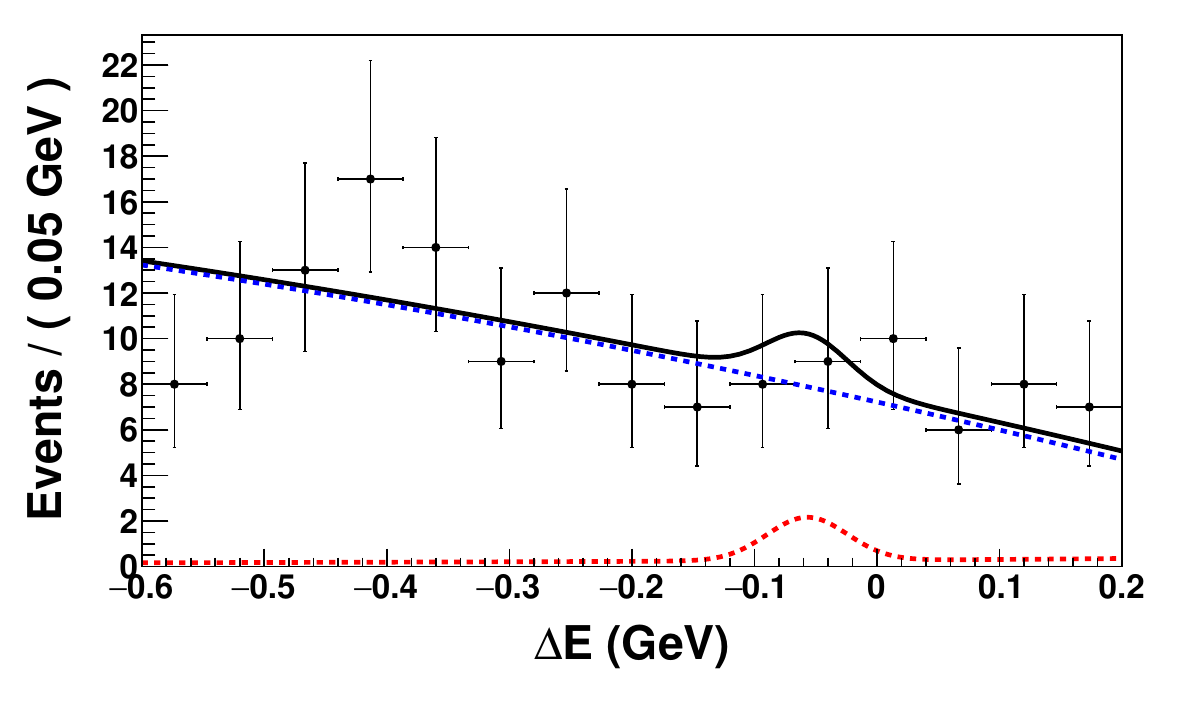}
\includegraphics[ height=4cm, width=5.9cm]{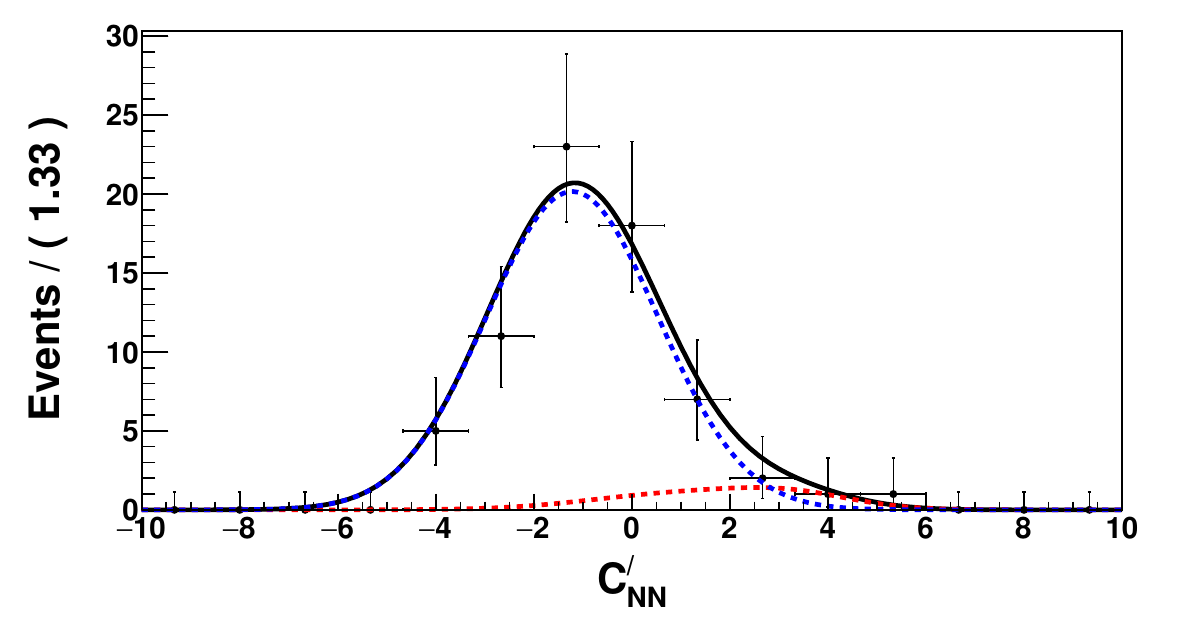}

\label{data_fit_23}
\centering
\includegraphics[ height=4cm, width=5.9cm]{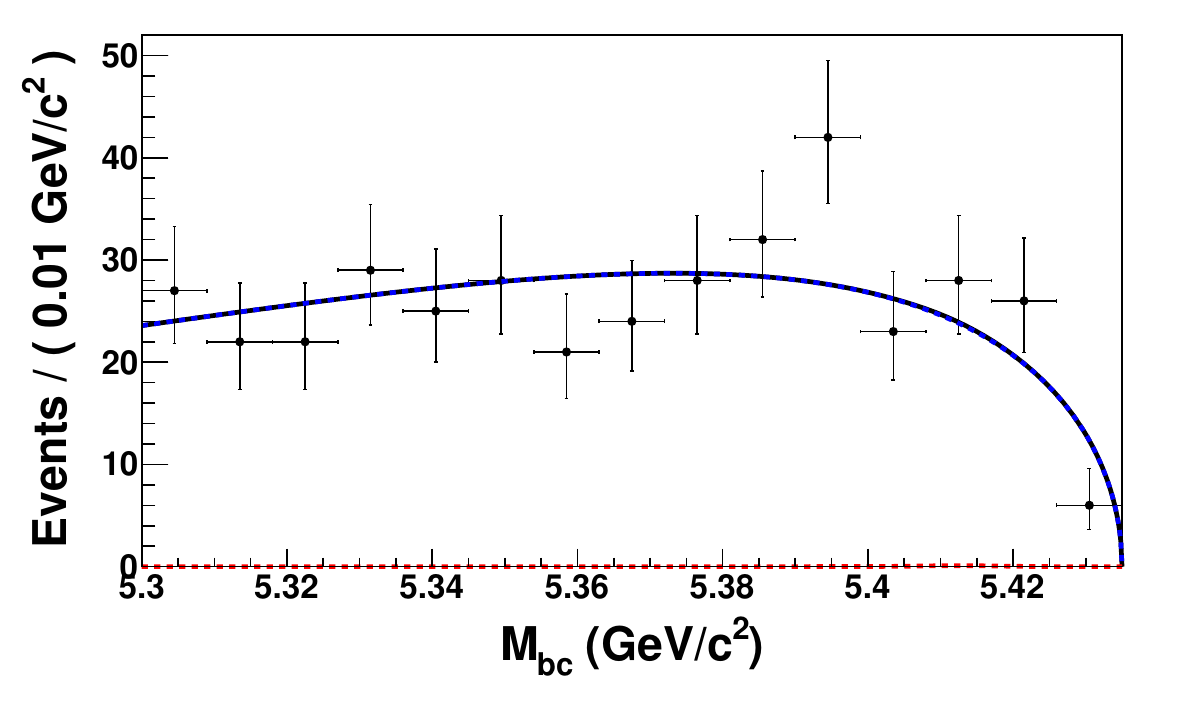}
\includegraphics[ height=4cm, width=5.9cm]{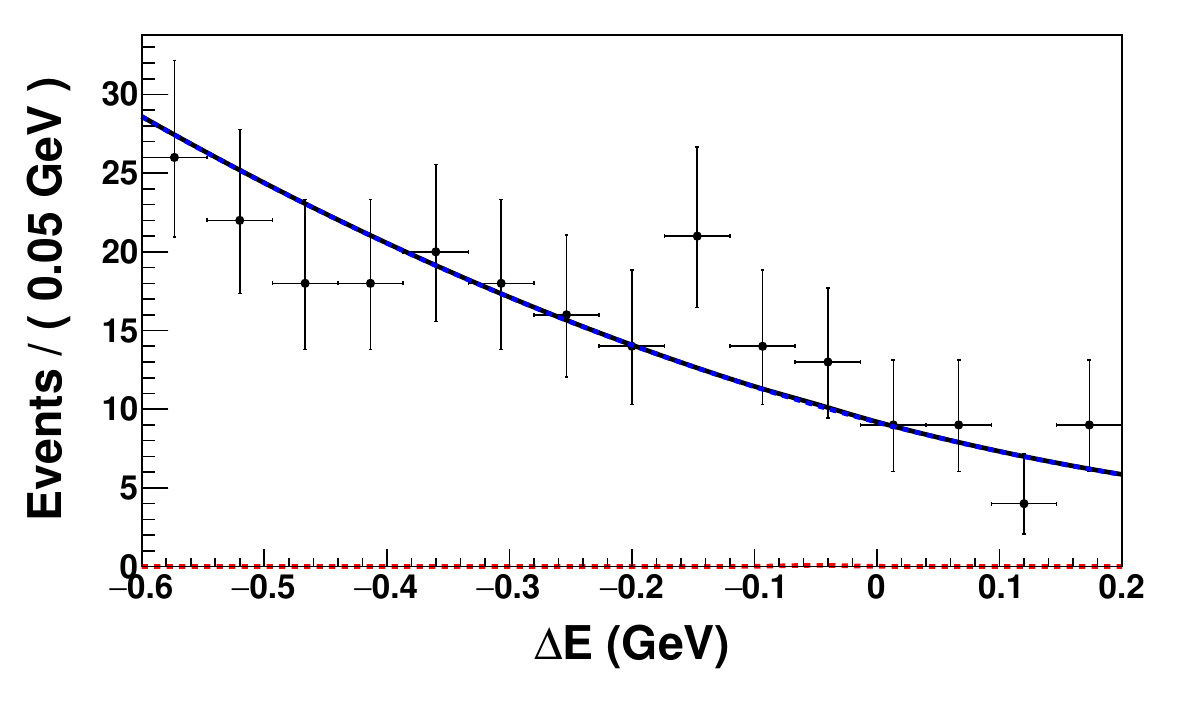}
\includegraphics[ height=4cm, width=5.9cm]{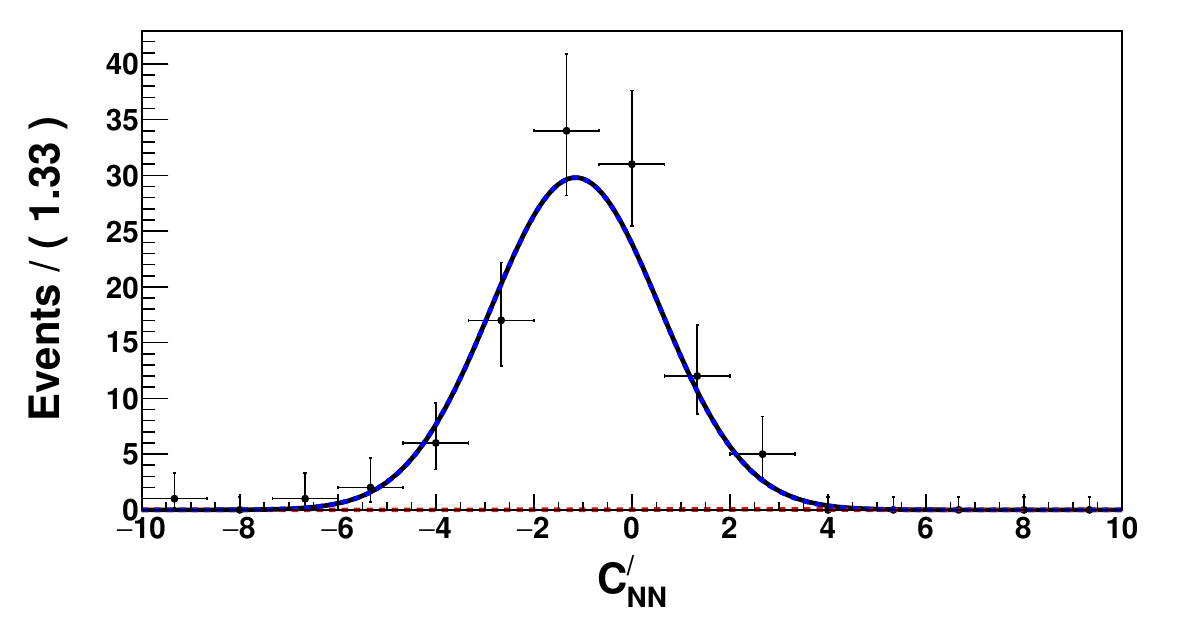}
\caption{Fit results for the $B_s^{0}\rightarrow\eta_{\gamma\gamma}\eta_{\gamma\gamma} $ (top), $B_s^{0}\rightarrow\eta_{\gamma\gamma}\eta_{3\pi}$ (middle) and $B_s^{0}\rightarrow\eta_{3\pi}\eta_{3\pi}$ (bottom) modes. The projections are shown for events inside a signal region in $M_{\rm bc}$, $\Delta E$ and $ \cal{C}'_{\mathrm{NN}}$, except for the variable plotted. The signal region is defined as  5.39 $<M_{\rm bc}<$ 5.43 GeV/$c^{2}$, -0.27  $< \Delta E <$ 0.12 GeV, and $ \cal{C}'_{\mathrm{NN}}>$ 2.91. Points with the error bars show the data, solid black curves show the total fit function, and dotted red (dashed blue) curves show the signal (background) contribution.}
\label{data_fit_33}
\end{figure*}

We extract signal yields for the three decay modes by performing an unbinned maximum likelihood fit to the variables $M_{\rm bc}$, $ \Delta E $, and $\cal{C}'_{\mathrm{NN}}$. We consider signal candidates originating only from $B_s^*\bar{B_s^*}$ production. The likelihood function is defined as

\begin{linenomath}
\begin{equation}
\mathcal{L}_{\rm fit} = e^{-\sum\limits_{j} n_j} \mathbf{\prod_{\textit{i}}^{\textit{N}}} \left(\sum\limits_{j} n_j P_j ( M_{\rm bc}^i, \Delta E^i, {\cal{C}'_{\mathrm{NN}}}^i)\right),
\end{equation}
\end{linenomath}
where $P_j ( M_{\rm bc}^i, \Delta E^i, {\cal{C}'_{\mathrm{NN}}}^i)$ is the probability density function (PDF) of the signal or background component (specified by index $j$), $n_j$ is the yield of this component, and $\textit N$ is the total number of events in the sample. The signal and background PDFs are determined from the respective MC samples, after applying all selection criteria. These PDFs used for modeling the signal, SCF, and continuum background are listed in Table \ref{tab:table1}. Correlations among the fit variables are found to be negligible. Thus, the three-dimensional PDF $P_j$ is expressed as the product of three one-dimensional PDFs, 

\begin{linenomath}
\begin{equation}
P_j \equiv P_j( M_{\rm bc})P_j (\Delta E) P_j(\cal{C}'_{\mathrm{NN}}). 
\end{equation}
\end{linenomath}The signal PDF parameters and the endpoint of the ARGUS function for the background $M_{bc}$ distribution are fixed to values obtained from an MC study, while all the other background PDF parameters are floated. The signal and background yields are floated in the fit, while the fraction of SCF events is fixed. The mean and standard deviation of the signal PDFs are corrected for small data-MC differences. These corrections are obtained by comparing the shapes of data and MC distributions for a control sample of $B^{0} \rightarrow \eta \eta$ decays, which have the same final-state particles.


\begin{table}[htb]
\caption{\label{tab:table2} Summary of fit results. The quoted uncertainties are statistical only.}
\resizebox{\columnwidth}{!}{\begin{tabular}{cccc}
\hline\hline
 Decay mode & $\eta_{\gamma\gamma}\eta_{\gamma\gamma}$ & $\eta_{\gamma\gamma}\eta_{3\pi}$ & $\eta_{3\pi}\eta_{3\pi}$ \\
\hline
Signal & $ 3.9\pm 3.2 $ & $ 7.7\pm 4.4 $ & $ 0.2\pm 5.9 $\\
Background & $ 882\pm 30 $ & $ 797\pm 28 $ & $ 1244\pm 36 $  \\
Efficiency $(\%)$ & $9.68\pm 0.09$ & $9.48\pm 0.09$ & $8.14\pm 0.08$\\
\hline 
$\mathcal{B}$ ($10^{-5}$) & $1.84\pm 1.38$ & $6.54\pm 3.27$ & $1.65\pm 9.90$ \\
\hline
Total $\mathcal{B}$  & & ($10.03\pm 10.52)\ \times 10^{-5}$ &\\
\hline
$\mathcal{B}$ UL at $90\%$ CL & & $14.3\times 10^{-5}$ & \\
\hline\hline

\end{tabular}}
\end{table}

From fitting the data, we obtain $3.9\pm 3.2$, $7.7\pm 4.4$, and $0.2\pm 5.9$ signal events for $\eta_{\gamma\gamma}\eta_{\gamma\gamma}$, $\eta_{\gamma\gamma}\eta_{3\pi}$, and $\eta_{3\pi}\eta_{3\pi}$ decay modes, respectively.
The common branching fraction by taking into account the signal yields of each decay mode (with index $\rm k$) is expressed as,

\begin{linenomath}
  \begin{equation}
\mathcal{B}(B^0_s \to \eta \eta) = \sum_{\rm k=1}^{3}\frac{N^{\rm sig}_{\rm k}}{N_{B_{s} \bar{B}_{s}} \times  \epsilon^{\rm rec}_{\rm k} \times \prod \mathcal{B}_{\eta \rm k}}
\end{equation}
\end{linenomath} where $N_{B_{s} \bar{B}_{s}}$ is the number of $B_{s} \bar{B}_s$ pairs; $\epsilon^{\rm rec}_{\rm k}$ and $N^{\rm{sig}}_{\rm k}$ is the signal selection efficiency obtained from MC simulation and number of signal events, respectively, for each of the decay modes; and $\prod \mathcal{B}_{\eta \rm k}$ is the product of the two $\eta$-decay branching fractions corresponding to each of the decay modes \cite{pdg2020}. The fit results are summarized in Table \ref{tab:table2}, and their projections in the signal regions are shown in Fig. \ref{data_fit_33}. 


The systematic uncertainties associated with the analysis are listed in Table \ref{tab:table3}. To investigate possible fit bias, we perform an ensemble test in which signal MC events are generated using EVTGEN and subsequently passed through a detector simulation based on GEANT3. The background events are generated from the corresponding PDFs used for fitting. For different numbers of input signal events, we generate and fit ensembles of $3000$ experiments each. From the pull distributions obtained from these ensemble tests, we observe an average fit bias of $6\%$, which we include as a source of systematic uncertainty.


The uncertainty due to PDF modeling is estimated from the variation in the signal yield while varying each fixed parameter by $\pm 1\sigma$. The uncertainty in the selection efficiency of the $\eta\rightarrow \gamma\gamma$ or $\pi^0\rightarrow \gamma\gamma$ decay is $3.0\%$, determined from a comparison of the data and signal MC selection efficiency ratios for large samples of $\eta\rightarrow\pi^+\pi^-\pi^0$ and $\eta\rightarrow 3\pi^0$ decays \cite{etapioneff}.
The systematic uncertainties due to the charged pion identification and tracking efficiency are $3.2\%$ and $2.1\%$ respectively, measured using control samples of $D^{*+}\rightarrow D^{0}\pi^{+}$, $D^{0} \rightarrow K^{-}\pi^{+} $, and $D^{*+}\rightarrow D^{0}\pi^{+}$, $D^{0} \rightarrow K_{S}^0\pi^{+}\pi^{-}$, $K_{S} \rightarrow \pi^{+} \pi^{-}$ decays. The systematic uncertainty due to the $\cal{C}'_{\mathrm{NN}}$ requirement is estimated by comparing the efficiencies in data and MC simulations of a large $B^0\rightarrow\eta\eta$ control sample. As mentioned above, the shapes of the signal PDFs are calibrated using the $B^0 \rightarrow \eta \eta$ control sample. The $M_{\rm bc}$ mean is found to be consistent within the statistical uncertainty, while the $\Delta E$ mean is shifted by 3 MeV. The systematic uncertainty associated with the small difference in $\cal{C}'_{\mathrm{NN}}$ mean and width as well as $M_{\rm bc}$ and $\Delta E$ width values are evaluated by varying these parameters by $\pm 3\sigma$, and is measured to be $3.1\%$. The uncertainty in the signal reconstruction efficiency due to MC statistics is $1.8\%$. We also assign a systematic uncertainty of $4.7\%$ due to the uncertainty in the prodution cross-section of $b\bar{b}$ events at $\Upsilon(5S)$ resonance. Systematic uncertainties of $0.5\%$ and $1.2\%$ are assigned due to the branching fractions of $\eta\rightarrow\gamma\gamma$ and $\eta\rightarrow\pi^+\pi^-\pi^0$ \cite{pdg2020}. 
Lastly, the uncertainty due to the fraction, $f_s$ is  $15.4\%$ \cite{pdg2020}.



\begin{table}[htb]
\caption{\label{tab:table3} Summary of systematic uncertainties. The uncertainties listed in the lower section of the table are external to the analysis.}
\begin{ruledtabular}
\begin{tabular}{c c}

Source & Value $(\%)$ \\
\hline
&\\
Fit bias & $\pm$6.0 \\

PDF modeling & 8.0 \\



$\eta\rightarrow\gamma\gamma$ selection efficiency & 3.0 \\

$\pi^0\rightarrow\gamma\gamma$ selection efficiency & 3.0 \\

Pion identification efficiency  & 3.2\\ 

Tracking efficiency  & 2.1 \\

$\cal{C}'_{\mathrm{NN}}$ requirement &  10.6\\

Calibration factors  & 3.1\\

MC statistics  & 1.8 \\

$L_{\rm int}$ & 1.3\\
&\\
\hline
& \\

$\sigma^{\Upsilon(5S)}_{b\bar{b}}$ & 4.7 \\

$\mathcal{B}$$(\eta\rightarrow\gamma\gamma)$ & 0.5 \\

$\mathcal{B}$$(\eta\rightarrow\pi^+\pi^-\pi^0)$ & 1.2 \\

 $f_s$ & 15.4 \\
&\\
\hline

Total & 22.8 \\ 

\end{tabular}
\end{ruledtabular}
\end{table}


Using the signal yields obtained from the fits of the three decay modes, the combined branching fraction for $\mathcal{B}(B^0_{s} \rightarrow \eta \eta)$ and $f_{s} \times \mathcal{B}(B^0_{s} \rightarrow \eta \eta)$ is calculated to be $(10.0 \pm 10.5 \pm 2.3) \times 10^{-5}$ and $(2.0 \pm 2.1 \pm 0.3) \times 10^{-5}$, respectively. The first uncertainty is statistical and the second is systematic. As we do not observe any significant signal yield, we use a Bayesian approach to set an upper limit(UL) on the branching fraction by integrating the combined likelihood function from 0 to 90\% of the total area under the curve for $\mathcal{B}$ $\geq\ 0$. The results from the  three $B_s^0\rightarrow\eta\eta$  decay modes are combined by adding the three individual log likelihoods as a function of branching fraction. Systematic uncertainties  are included by convolving the combined likelihood curve with a Gaussian function of width equal to the total systematic uncertainty mentioned in Table III.
We obtain a $90\%$ CL UL of $14.3\times 10^{-5}$ on the branching fraction. 
In addition, the $90\%$ CL UL  on the product $f_s \times \mathcal{B}(B^0_{s} \rightarrow \eta \eta)$ is estimated to be less than $2.9\times 10^{-5}$. The total fractional systematic uncertainty associated with $\mathcal{B}(B^0_{s} \rightarrow \eta \eta)$ and $f_{s} \times \mathcal{B}(B^0_{s} \rightarrow \eta \eta)$ is $22.8\%$ and $16.8\%$, respectively.
These results are summarized in Table IV.
\begin{table}[htb]
\caption{Summary of results on branching fractions and UL for $\mathcal{B}(B^0_{s} \rightarrow \eta \eta)$ and $f_{s} \times \mathcal{B}(B^0_{s} \rightarrow \eta \eta)$.}
\begin{ruledtabular}
	\begin{tabular}{c c}
		Quantity & Value \\
		\hline
		         &       \\
		$\mathcal{B}(B^0_{s} \rightarrow \eta \eta)$ & 
		                                             $(10.0 \pm 10.5 \pm 2.3) \times 10^{-5}$ \\
		                                             & $< 14.3 \times 10^{-5}$ @ $90\%$
		                                             CL \\
		$f_{s} \times \mathcal{B}(B^0_{s} \rightarrow \eta \eta)$                                       
		                                             & $(2.0 \pm 2.1 \pm 0.3) \times 10^{-5}$ \\ 
		                                             & $< 2.9 \times 10^{-5}$ @ $90\%$ CL \\
	\end{tabular}
\end{ruledtabular}  
\end{table}                                           
	





In summary, we have searched for the decay $B_s^0\rightarrow\eta\eta$ using the complete $\Upsilon(5S)$ dataset from the Belle experiment. We do not observe any statistically  significant signal for the decay and set a $90\%$ confidence level upper limit of $14.3\times 10^{-5}$ on its branching fraction. This is an improvement by about an order of magnitude over the previous limit \cite{l3}.

\begin{center}
{\bf ACKNOWLEDGEMENTS}
\end{center}

We thank the KEKB group for the excellent operation of the
accelerator; the KEK cryogenics group for the efficient
operation of the solenoid; and the KEK computer group, and the Pacific Northwest National
Laboratory (PNNL) Environmental Molecular Sciences Laboratory (EMSL)
computing group for strong computing support; and the National
Institute of Informatics, and Science Information NETwork 5 (SINET5) for
valuable network support.  We acknowledge support from
the Ministry of Education, Culture, Sports, Science, and
Technology (MEXT) of Japan, the Japan Society for the 
Promotion of Science (JSPS), and the Tau-Lepton Physics 
Research Center of Nagoya University; 
the Australian Research Council including grants
DP180102629, 
DP170102389, 
DP170102204, 
DP150103061, 
FT130100303; 
Austrian Science Fund under Grant No.~P 26794-N20;
the National Natural Science Foundation of China under Contracts
No.~11435013,  
No.~11475187,  
No.~11521505,  
No.~11575017,  
No.~11675166,  
No.~11705209;  
Key Research Program of Frontier Sciences, Chinese Academy of Sciences (CAS), Grant No.~QYZDJ-SSW-SLH011; 
the  CAS Center for Excellence in Particle Physics (CCEPP); 
the Shanghai Pujiang Program under Grant No.~18PJ1401000;  
the Ministry of Education, Youth and Sports of the Czech
Republic under Contract No.~LTT17020;
the Carl Zeiss Foundation, the Deutsche Forschungsgemeinschaft, the
Excellence Cluster Universe, and the VolkswagenStiftung;
the Department of Science and Technology of India; 
the Istituto Nazionale di Fisica Nucleare of Italy; 
National Research Foundation (NRF) of Korea Grants
No.~2015H1A2A1033649, No.~2016R1D1A1B01010135, No.~2016K1A3A7A09005
603, No.~2016R1D1A1B02012900, No.~2018R1A2B3003 643,
No.~2018R1A6A1A06024970, No.~2018R1D1 A1B07047294; Radiation Science Research Institute, Foreign Large-size Research Facility Application Supporting project, the Global Science Experimental Data Hub Center of the Korea Institute of Science and Technology Information and KREONET/GLORIAD;
the Polish Ministry of Science and Higher Education and 
the National Science Center;
the Grant of the Russian Federation Government, Agreement No.~14.W03.31.0026; 
the Slovenian Research Agency;
Ikerbasque, Basque Foundation for Science, Spain;
the Swiss National Science Foundation; 
the Ministry of Education and the Ministry of Science and Technology of Taiwan;
and the United States Department of Energy and the National Science Foundation.



\begin{thebibliography}{99}


%
%
%

\bibitem{scet} 
 A.~R.~Williamson and J.~Zupan,
  Phys.\ Rev.\ D {\bf 74}, 014003 (2006)
  [Erratum: Phys.\ Rev.\ D {\bf 74}, 03901 (2006)].

\bibitem{pqcd} 
  A.~Ali {\it et al.},
  Phys.\ Rev.\ D {\bf 76}, 074018 (2007).
  
\bibitem{qcdf} 
M.~Beneke and M.~Neubert,
Nucl.\ Phys.\ B {\bf 675} (2003) 333-415.

\bibitem{l3}
M. Acciarri {\it et al.} (L3 Collaboration), Phys.\ Lett.\ B {\bf 363}, 127 (1995).

\bibitem{detector1} 
  A.~Abashian {\it et al.} (Belle Collaboration),
  Nucl.\ Instrum.\ Methods Phys. Res., Sect \ A {\bf 479}, 117 (2002).

\bibitem{detector2} 
  J.~Brodzicka {\it et al.} (Belle Collaboration),
  PTEP {\bf 2012}, 04D001 (2012)

\bibitem{Sevda} S. Esen {\it et al.} (Belle Collaboration), Phys.\ Rev.\ D {\bf 87}, 031101(R) (2013).

\bibitem{pdg2020} P. A. Zyla {\it et al.} (Particle Data Group), Prog. Theor. Exp. Phys. \textbf{2020}, 083C01 (2020).

\bibitem{evtgen}
D. J. Lange, Nucl. Instrum. Methods Phys. Res., Sect. A
{\bf 462}, 152 (2001).

\bibitem{geant3}
R. Brun {\it et al.} GEANT 3.21. Report No. CERN
DD/EE/84-1 (1984).

\bibitem{CBall} T.~Skwarnicki, Ph.D. thesis, Institute for Nuclear Physics, Krakow, DESY Internal Report, DESY F31-86-02 (1986).

\bibitem{Argus} H. Albrecht {\it et al.} (ARGUS Collaboration), Phys.\ Lett.\ B {\bf 241}, 278 (1990).

\bibitem{nn}
M. Feindt and U. Kerzel, Nucl. Instrum. Methods Phys.
Res., Sect. A {\bf 559}, 190 (2006).

\bibitem{ksfw}
G. C. Fox and S. Wolfram, Phys. Rev. Lett. {\bf 41}, 1581
(1978); S. H. Lee {\it et al.} (Belle Collaboration), Phys. Rev.
Lett. {\bf 91}, 261801 (2003).



\bibitem{Bfactories} A. J. Bevan, B. Golob, Th. Mannel, S. Prell, and B. D. Yabsley, Eur. Phys. J. C {\bf 74}, 3026 (2014).

\bibitem{etapioneff} M. C. Chang {\it et al.} (Belle Collaboration), Phys. Rev. D {\bf 85}, 091102(R) (2012).


\end{thebibliography}
\end{document}